\documentclass[aps,showpacs,showkeys]{revtex4}
\usepackage{amsmath,amsthm,amssymb,epsfig,alltt}

\begin{document}

\def\a{\alpha}
\def\b{\beta}
\def\d{{\delta}}
\def\l{\lambda}
\def\e{\epsilon}
\def\p{\partial}
\def\m{\mu}
\def\n{\nu}
\def\t{\tau}
\def\th{\theta}
\def\s{\sigma}
\def\g{\gamma}
\def\o{\omega}
\def\r{\rho}
\def\half{\frac{1}{2}}
\def\hatt{{\hat t}}
\def\hatx{{\hat x}}
\def\hatp{{\hat p}}
\def\hatX{{\hat X}}
\def\hatY{{\hat Y}}
\def\hatP{{\hat P}}
\def\haty{{\hat y}}
\def\whatX{{\widehat{X}}}
\def\whata{{\widehat{\alpha}}}
\def\whatb{{\widehat{\beta}}}
\def\whatV{{\widehat{V}}}
\def\hatth{{\hat \theta}}
\def\hatta{{\hat \tau}}
\def\hatrh{{\hat \rho}}
\def\hatva{{\hat \varphi}}
\def\barx{{\bar x}}
\def\bary{{\bar y}}
\def\barz{{\bar z}}
\def\baro{{\bar \omega}}
\def\barpsi{{\bar \psi}}
\def\sp{\sigma^\prime}
\def\nn{\nonumber}
\def\cb{{\cal B}}
\def\2pap{2\pi\alpha^\prime}
\def\wideA{\widehat{A}}
\def\wideF{\widehat{F}}
\def\beq{\begin{eqnarray}}
 \def\eeq{\end{eqnarray}}
 \def\4pap{4\pi\a^\prime}
 \def\xp{{x^\prime}}
 \def\sp{{\s^\prime}}
 \def\ap{{\a^\prime}}
 \def\tp{{\t^\prime}}
 \def\zp{{z^\prime}}
 \def\xpp{x^{\prime\prime}}
 \def\xppp{x^{\prime\prime\prime}}
 \def\barxp{{\bar x}^\prime}
 \def\barxpp{{\bar x}^{\prime\prime}}
 \def\barxppp{{\bar x}^{\prime\prime\prime}}
 \def\barchi{{\bar \chi}}
 \def\baro{{\bar \omega}}
 \def\bpsi{{\bar \psi}}
 \def\barg{{\bar g}}
 \def\barz{{\bar z}}
 \def\bareta{{\bar \eta}}
 \def\ta{{\tilde \a}}
 \def\tb{{\tilde \b}}
 \def\tc{{\tilde c}}
 \def\tz{{\tilde z}}
 \def\tJ{{\tilde J}}
 \def\tpsi{\tilde{\psi}}
 \def\tal{{\tilde \alpha}}
 \def\tbe{{\tilde \beta}}
 \def\tga{{\tilde \gamma}}
 \def\tchi{{\tilde{\chi}}}
 \def\barth{{\bar \theta}}
 \def\bareta{{\bar \eta}}
 \def\barom{{\bar \omega}}
 \def\bole{{\boldsymbol \epsilon}}
 \def\bolth{{\boldsymbol \theta}}
 \def\bomega{{\boldsymbol \omega}}
 \def\bolmu{{\boldsymbol \mu}}
 \def\bola{{\boldsymbol \alpha}}
 \def\bolb{{\boldsymbol \beta}}

\setcounter{page}{1}
\title[]{The Tomonaga-Luttinger Liquid with Quantum Impurity Revisited: \\
Critical Line and Phase Diagram}

\author{Taejin Lee}
\affiliation{
Department of Physics, Kangwon National University, Chuncheon  24341 
Korea}

\email{taejin@kangwon.ac.kr}

\begin{abstract}
We revisit the $(1+1)$ dimensional field theoretical model, which describes the Tomonaga-Luttinger liquid (TLL), interacting with a static impurity at the origin of the half line. Applying the Fermi-Bose equivalence and finite conformal transformations only, we map the model onto the Schmid model. 
Some details of the bosonization procedure 
have been given. The critical line and the phase diagram of the model follow from the renormalization group
analysis of the Schmid model. The obtained critical line of the model is a hyperbola in the parameter space of the two couplings of the TLL. 
\end{abstract}




\maketitle

\section{Introduction}

The Tomonaga-Luttinger liquid (TLL) model \cite{Tomonaga, Luttinger, Haldane} has received  much attention from both theorists and experimentalists for decades, since it is the most essential component of the $(1+1)$ dimensional field theories, 
which describe the strongly interacting, non-Fermi liquids in one dimension. 
The TLL model has been discussed in various contexts in the literature, such as one dimensional quantum
wires \cite{kane;1992b, kane1992r, Kane:1992a, Furusaki93}, Josephson junctions arrays \cite{larkin, fazio, sodano, Giuliano08}, Kondo effects \cite{Affleck:1990by, Affleck:1990iv, Furusaki94}, 
carbon nanotube \cite{Egger} 
and edge states of quantum Hall effects \cite{kane2, fendley1995a, chang}. The 
TLL model in the presence of an impurity is formulated as a one dimensional fermion theory 
with four fermi interactions on a half line and 
a tunneling interaction through an impurity at the origin. The main tool to analyze the model is the 
Fermi-Bose equivalence \cite{Mattis65, Mandelstam, coleman75}, which is also called as fermionization or bosonization in the literature \cite{stone94, delft, gogolin, Hassel, Lee2005, Leeklein}. Applying the 
Fermi-Bose equivalence to the fermion model, one obtains the boson form of the TLL model, which is characterized by the two parameters, termed as Tomonaga-Luttinger (TL) parameters \cite{voit, senechal} . 
However, these TL parameters seem to be different from those we would obtain by extending the 
work of Coleman on the equivalence between the massive Thirring model and the sine-Gordon model. 
The TLL model has two types of the four fermi interactions. If one of the four fermi interaction term, corresponding to the forward scattering, is turned off, the TLL model reduces to the Thirring model. 
Then we can identify the TL parameters in the case of this reduced model by employing the equivalence between the Thirring model and the sine-Gordon model in ref.\cite{coleman75}. (To be precise, the boundary sine-Gordon model for the case of the TLL model.)
These TL parameters differ from those given in the literature. There seems to be some discrepancy.   

In this paper we discuss the application of the Fermi-Bose equivalence with some rigor along the line of Coleman's work on the Thirring model. We shall briefly review of the Fermi-Bose equivalence to identify the 
setting where the Fermi-Bose equivalence is applicable. Then we recast TLL model into the setting to bosonize
the action by redefining the spatial and temporal worldsheet coordinates. Once the action is bosonized, 
it is rewritten in terms of the non-chiral boson field instead of chiral boson fields. 
The reason is that the four fermi interaction terms 
contain the time derivatives of the boson field, which require redefining the canonical momentum field.
Even if the bosonized terms are written as spatial derivatives of the chiral boson fields, it should be noted that they could be equally written also as time derivatives of the chiral boson fields. 
By only applying the finite conformal transformations, which the theory allows, we transcribe the boson model 
into the Schmid model \cite{schmid, guinea, fisher;1985}. 
The Schmid model has already appeared in the works of Kane and Fisher \cite{kane;1992b, kane1992r, Kane:1992a} 
on their study of transport in a one-channel Luttinger liquid.
However, since their discussions began with the bosonized action of the TLL model, the 
precise relations between the fermion TLL model and the Schmid model has not been given. 
The purpose of this work is to establish the equivalence of the TLL model and the Schmid model and to identify the precise relation between the couplings of the TLL model and those of the Schmid model. 
Along this line we may be able to determine the TL parameters uniquely.

\section{The Fermi-Bose Equivalence}
The Fermi-Bose equivalence is the one of the most powerful tools to analyze  $(1+1)$ dimensional interacting theories. Here we review briefly the equivalence with some details before applying it to the TLL model.
Because the Fermi-Bose equivalence plays a crucial role in the analysis and requires certain conditions to be 
fulfilled.
We consider the following action of a free boson field on a cylindrical space-time with Euclidean signature
\beq \label{bosonaction}
S = \frac{\a}{4\pi} \int_{0}^\infty d\t  \int_{-\pi}^{\pi}d\s \left(\p_\t X \p_\t X + \p_\s X \p_\s X \right)
\eeq
where the parameter $\a$ in Eq.(\ref{bosonaction}) corresponds to the inverse of the Regge slope $\ap = 1/\a$, in string theory and the TL parameter of the bosonized TTL model in condensed matter physics.
Since the boson field is defined on a cylindrical space-time, we may expand it as 
\beq \label{normal}
X(\s) = Rm\s + \sum_{n \in Z} X_n e^{-in\s}.
\eeq
Here $R$ is the radius of the circle on which the boson field is compactified and the integer $m$ is the 
winding number. In terms of the normal mode expansions Eq.(\ref{normal}), the action of the boson field 
may be written as 
\begin{subequations}
\beq
S &=& \int^\infty_0 d\t L, \\
L &=& \a \left[ \half R^2 m^2 +  \half \sum_{n \in Z} \left(\dot{X}_n                                    
\dot{X}_{-n} + n^2 X_n X_{-n}\right) \right].                                                   
\eeq     
\end{subequations}
Defining the canonical momenta 
\beq
\quad P(\s) = \sum_{n \in Z} P_n e^{in\s},~~~P_0 = p,~ \text{has an eigenvalue},~ n/R,  \,\,\, n\in Z ,
\eeq
we obtain the Euclidean Hamiltonian as
\beq                                                                                     
P_n &=& i\frac{\p L}{\p \dot{X}_n} = i\a \dot{X}_{-n}, \nn\\                     
H &=& i \sum_{n \in Z} P_n \dot X_n +L \nn\\
&=& \frac{\a}{2} R^2 m^2 +\half \sum_{n\in Z}\left(\frac{1}{\a} P_n P_{-n} + \a n^2 X_n X_{-n} \right) .  
\eeq                                                                                     
The canonical commutation relations are
\beq
[P_n, X_n ] = \frac{1}{i}, ~~ n \in Z .
\eeq

It is convenient to define the left moving and right moving oscillatory operators as usual
\beq
\a_n = \half\left(lP_{-n} + \frac{2n}{il} X_n \right), \quad
\tilde{\a}_n = \half \left(l P_n + \frac{2n}{i l} X_{-n} \right),
\eeq 
which satisfy the following commutation relations
\beq
[\a_n,\a_m] = [\tilde\a_n, \tilde\a_m] = n \delta(n+m), ~~~ n, m \in Z
\eeq
where $l = \sqrt{2/\a}$. 
Accordingly the Hamiltonian is given by
\beq
H = \frac{\a}{2}R^2m^2 + \frac{1}{2\a} p^2 + \sum_{n=1} n \left(a^\dagger_n a_n
+ \tilde{a}^\dagger_n \tilde{a}_n \right) + \sum n
\eeq
where 
\begin{subequations}
\beq
\a_n &=& \sqrt{n} a_n, \quad \a_{-n} = \sqrt{n} a^\dagger_n, \quad
\tal_n = \sqrt{n} \ta_n, \quad \tal_{-n} = \sqrt{n} \ta^\dagger_n, \quad n>0 \\ 
\left[a_n, a^\dagger_m \right] &=& \delta_{nm}, \quad 
\left[\tilde a_n, \tilde a^\dagger_m \right] = \delta_{nm}, \quad 
\a_0 = \tilde{\a}_0 = \half 
l p. 
\eeq 
\end{subequations}

The Euclidean Hamiltonian defines the time evolution of the operators                                                     
\begin{subequations}
\beq                                                             
\a_n e^{-in\s} &\rightarrow &  e^{\t H} \a_n e^{-in\s} e^{-\t H} = \a_n     
e^{-n(\t+i\s)}, \\
\tilde\a_n e^{in\s}&\rightarrow &  e^{\t H} \tilde\a_n e^{in\s} e^{-\t H} = \a_n     
e^{-n(\t-i\s)}.                                                 
\eeq
\end{subequations}
It leads us to decompose the boson field $X$ into two chiral boson fields $X_L$ and $X_R$                                                             as \cite{Giveon}
\begin{subequations}
\beq
X &=& X_L + X_R, \\
X_L(\t+i\s) &=& \frac{1}{\sqrt{2\a}} x_L - \frac{i}{\sqrt{2\a}} p_L(\t+i\s) + \frac{i}{\sqrt{2\a}} \sum_{n\not=0}
\frac{\a_n}{n} e^{-n(\t+i\s)}, \\
X_R(\t-i\s) &=& \frac{1}{\sqrt{2\a}} x_R - \frac{i}{\sqrt{2\a}} p_R(\t-i\s) + \frac{i}{\sqrt{2\a}} \sum_{n\not=0}
\frac{\tilde\a_n}{n} e^{-n(\t-i\s)},
\eeq
\end{subequations}
where $x = \frac{1}{\sqrt{2\a}}\left(x_L + x_R\right)$, $[x_L,p_L] = [x_R,p_R] = i$, and
\beq
p_L = \frac{1}{\sqrt{2}} \left(\frac{n}{\sqrt{\a}R} + \sqrt{\a}R m \right), ~~
p_R = \frac{1}{\sqrt{2}} \left(\frac{n}{\sqrt{\a}R} - \sqrt{\a}R m \right).
\eeq
The zero mode component of the Hamiltonian is written as
\beq
H_0 = \frac{1}{2} \left(p_L^2 + p_R^2 \right). 
\eeq

Now we come to the crux of the Fermi-Bose equivalence.
Let us write the fermion operators as 
\begin{subequations}
\beq
\psi_L &=& : e^{-i\b X_L}:\nn\\
&=& : \exp\left[ - \frac{i\b}{\sqrt{2\a}}\left( x_L -i p_L(\t+i\s) + i \sum_{n\not=0}\frac{\a_n}{n} e^{-n(\t+i\s)} \right) \right]: ,  \label{psil}\\
\psi_R &=& :e^{i \b X_R}: \nn\\
&=& : \exp\left[ \frac{i\b}{\sqrt{2\a}}\left( x_R -i p_R(\t-i\s) + i \sum_{n\not=0}\frac{\tilde\a_n}{n} e^{-n(\t-i\s)} \right) \right]: \label{psir}
\eeq
\end{subequations}
where $\b$ is a constant parameter to be fixed shortly. Let us call these operators as Mattis-Mandelstam (MM) operators. 
Under the translation $\s \rightarrow \s+ 2\pi$ the fermion operators must be periodic or anti-periodic: 
\beq
\psi_L(\t, \s+2\pi) &=&  \pm \psi_L(\t, \s) , ~~~ \psi_R(\t, \s+2\pi) =  \pm \psi_R(\t, \s) .
\eeq
These conditions may be written as 
\begin{subequations}
\beq
\exp\left[-i\b \sqrt{\frac{2}{\a}} p_L \, \pi \right] &=& \pm 1, \label{cond1}\\
\exp\left[-i\b \sqrt{\frac{2}{\a}} p_R \, \pi \right] &=& \pm 1 \label{cond2}.
\eeq
\end{subequations}
We may also require the Dirac mass operator to be periodic under the translation, 
$X \rightarrow X + 2\pi R$,
\beq
\psi^\dag_L \psi_R =: \exp\left(i \b (X_L+X_R) \right): = :\exp\left( i\b X \right): 
\eeq
It yields 
\beq
\b = 1/R.
\eeq
Then the conditions Eqs.(\ref{cond1}, \ref{cond2}) are satisfied as
\beq
\b \sqrt{\frac{2}{\a}} p_L = \frac{n}{\a R^2} + m = \text{integer}, ~~~ 
\b \sqrt{\frac{2}{\a}} p_R = \frac{n}{\a R^2} - m = \text{integer}.
\eeq
It imposes a constraint condition on $R$
\beq
R^2 = \frac{1}{k \a}, ~~~ k \in Z .
\eeq
  
In order to fix the integer $k$, let us consider the fermion Green's function. 
Using Eqs.(\ref{psil}, \ref{psir}), we obtain for $\t > \t^\prime$,
\begin{subequations}
\beq
\langle 0 \vert \psi_L(\t,\s) \psi^\dag_L(\t^\prime, \s^\prime) \vert 0 \rangle 
&=& \left[\frac{\sqrt{z z^\prime}}{z-z^\prime}\right]^{\frac{\b^2}{2\a}} , \\
\langle 0 \vert \psi_R(\t,\s) \psi^\dag_R(\t^\prime, \s^\prime) \vert 0 \rangle 
&=& \left[\frac{\sqrt{\barz \barz^\prime}}{\barz-\barz^\prime}\right]^{\frac{\b^2}{2\a}},
\eeq
\end{subequations}
where $z = e^{\t+ i\s}$ and  $\barz = e^{\t-i\s}$. 
It reduces precisely to the fermion Green's function in the Neveu-Schwarz (NS) 
sector on a cylindrical surface, if 
\beq
\frac{\b^2}{2\a}=\frac{k}{2}=1.
\eeq
Thus, the MM operators Eqs.(\ref{psil}, \ref{psir}) are equivalent to the fermion field operators when
\beq \label{radius}
\b = \sqrt{2\a}, ~~~ R = 1/\sqrt{2\a} .
\eeq
Without loss of generality, we may choose $\a=1$; $R= 1/\sqrt{2}$: The fermion operators are defined as
\begin{subequations}
\beq
\psi_L &=& : e^{-i\sqrt{2} X_L}:
= : \exp\left[- i\left( x_L -i p_L(\t+i\s) + i \sum_{n\not=0}\frac{\a_n}{n} e^{-n(\t+i\s)} \right) \right]: , \label{fermionl} \\
\psi_R &=& :e^{i \sqrt{2} X_R}:
= : \exp\left[ i\left( x_R -i p_R(\t-i\s) + i \sum_{n\not=0}\frac{\tilde\a_n}{n} e^{-n(\t-i\s)} \right) \right]:, \label{fermionr}
\eeq
\end{subequations}
and they satisfy 
\beq
\{\psi^{ \dagger}_L(\s), \psi_L(\s^\prime)\} =  2\pi \d(\s-\s^\prime) , \quad 
\{\psi^{ \dagger}_R(\s), \psi_R(\s^\prime)\} =  2\pi \d(\s-\s^\prime). \label{antib}
\eeq

The Fermi-Bose equivalence is completed by introducing the Klein factors \cite{Leeklein}, 
\beq \label{fermionoperators}
\psi_L = : e^{-\frac{\pi}{2}i(p_L+p_R)} e^{-\sqrt{2}i X_L}:, ~~~
\psi_R = : e^{-\frac{\pi}{2}i(p_L+p_R)} e^{\sqrt{2}i X_R}: , 
\eeq
which ensure the anti-commutation relations between fermion operators
\begin{subequations}
\beq
\{\psi_L(\s), \psi_R(\s^\prime)\} &=& 0 , \quad 
\{\psi^{ \dagger}_L(\s), \psi_R(\s^\prime)\} = 0, \\ 
\{\psi_L(\s), \psi^\dag_R(\s^\prime)\} &=& 0, 
\quad \{\psi^\dag_L(\s), \psi^\dag_R(\s^\prime)\} = 0 . \label{antic}
\eeq
\end{subequations}
With the condition Eq.(\ref{radius}), the fermion field operators Eq.(\ref{fermionoperators}) are anti-periodic 
under the translation $\s \rightarrow \s + 2\pi$. It implies that we select the NS sector. 
It is consistent with the finite temperature field theory. 
The Fermi-Bose equivalence may be expressed in terms of the free field actions as 
\beq
S_B = \frac{1}{4\pi} \int^{\infty}_{0} d\t \int^{\pi}_{-\pi} d\s \,\left[\p X \cdot \p X \right]
\Longleftrightarrow S_F = \frac{1}{2\pi} \int^{\infty}_{0} d\t \int^{\pi}_{-\pi} d\s \,
\left[ \barpsi \g \cdot \p \psi \right] . \label{equivalence}
\eeq
The Fermi-Bose equivalence relations for the fermion bilinear operators follow from Eqs.(\ref{fermionl}, \ref{fermionr})
\begin{subequations}
\beq
:\psi^\dag_L \psi_L: &=& \sqrt{2} i \p_\t X_L = \sqrt{2}\p_\s X_L 
= \frac{i}{\sqrt{2}} \left(\p_\t - i\p_\s \right) X_L = \frac{i}{\sqrt{2}} \left(\p_\t - i\p_\s \right) X ,
\label{bilinear1} \\
:\psi^\dag_R \psi_R: &=&  -\sqrt{2} i \p_\t X_R = \sqrt{2}\p_\s X_R 
= -\frac{i}{\sqrt{2}} \left(\p_\t + i\p_\s \right) X_R = -\frac{i}{\sqrt{2}} 
\left(\p_\t + i\p_\s \right) X . \label{bilinear2}
\eeq
\end{subequations}
Since the Fermi-Bose equivalence works only in this particular setting, 
when we apply the Fermi-Bose equivalence to a theory, it is appropriate to recast the theory 
into this setting. We note also that the Fermi-Bose equivalence may not map the Hamiltonian of the 
fermion theory directly onto the Hamiltonian of the corresponding boson theory in the presence of interactions 
and vice versa.

\section{Tomonaga-Luttinger Liquid with Quantum Impurity}

The Tomonaga-Luttinger Liquid, interacting with an impurity at the origin of the half line is described by the 
following Euclidean fermion action
\beq
S &=& \frac{1}{2\pi} \int_0^\infty dx \int^{\b_T/2}_{-\b_T/2} dt \,\, \Biggl\{
\barpsi \left(\bar\g^0 \p_t + v \bar\g^1 \p_x \right) \psi + 
\bar g_2 \left(\psi^\dag_L \psi_L \psi^\dag_R \psi_R \right)
+ \frac{\bar g_4}{2} \left[ \left(\psi^\dag_L \psi_L \right)^2 + \left(\psi^\dag_R \psi_R \right)^2 \right]
\Biggr\} \nn\\
&& + \frac{1}{2\pi} \int^{\b_T/2}_{-\b_T/2} dt \, \bar \l \left(\psi^\dag_L \psi_R + 
\psi^\dag_R \psi_L  \right)\Bigl\vert_{x=0} ,
\eeq
where $\b_T = 1/T$. In order to apply the Fermi-Bose equivalence, we define new space-time variables, $\t$ and $\s$
\beq
x = \frac{v \b_T}{2\pi} \t, ~~~ t = \frac{\b_T}{2\pi} \s .
\eeq
At the same time we scale $\psi \rightarrow \psi/\sqrt{v}$, $\psi^\dag \rightarrow \psi^\dag/\sqrt{v}$. 
(It corresponds to a finite wave function renormalization.)
It brings us to the setting, which we can apply the Fermi-Bose equivalence Eq.(\ref{fermionl}, \ref{fermionr}, \ref{equivalence}), discussed in the previous section
\beq
S &=& \frac{1}{2\pi} \int_0^\infty d\t \int^{\pi}_{-\pi} d\s \,\, \Biggl\{
\barpsi \left(\g^0 \p_\t+ \g^1 \p_\s \right) \psi + g_2 \left(\psi^\dag_L \psi_L\right)
\left(\psi^\dag_R \psi_R \right) 
+ \frac{g_4}{2} \left[ \left(\psi^\dag_L \psi_L \right)^2 + \left(\psi^\dag_R \psi_R \right)^2 \right]
\Biggr\} \nn\\
&& + \frac{1}{2\pi} \int^{\pi}_{-\pi} d\s \,  \l \left( \psi^\dag_L \psi_R + 
\psi^\dag_R \psi_L  \right)\Bigl\vert_{\t=0} 
\eeq
where 
\beq
\g^0 = \bar \g^1, ~~~\g^1 = \bar \g^0 , ~~~ g_2 = \frac{\b_T^2}{4\pi^2 v} \bar g_2, ~~~
g_4 = \frac{\b_T^2}{4\pi^2 v} \bar g_4, ~~~ \l = \frac{\b_T}{2\pi v} \bar \l \, \, .
\eeq
Note that we work on the TLL theory at finite temperature. The periodic Euclidean time $t$ serves 
as the periodic world sheet coordinate $\s$ and the spatial coordinate $x$ plays the role of the 
world sheet coordinate $\t$ in string theory.  

The free field part of the fermion Lagrangian is written in terms of the boson field as 
\beq \label{free}
\frac{1}{2\pi} \bar \psi \g \cdot \p \psi = \frac{1}{4\pi} \p \phi \cdot \p \phi . 
\eeq
Making use of the Fermi-Bose equivalence for the fermion bilinear operators Eqs.(\ref{bilinear1}, \ref{bilinear2}),
we find that the second term in the fermion action may be written as the free boson field action  
\beq
:\left(\psi^\dag_L \psi_L\right)\left(\psi^\dag_R \psi_R\right) :
= \frac{1}{2} \left(\p_\t - i\p_\s \right) \phi \left(\p_\t + i\p_\s \right) \phi = 
\frac{1}{2} \p \phi \cdot \p \phi . \label{second}
\eeq
By the same identities of the Fermi-Bose equivalence Eqs.(\ref{bilinear1}, \ref{bilinear2}),
we can also transcribe the third term of the fermion action into the following boson bilinear term as 
\beq
:\left(\psi^\dag_L \psi_L\right)^2:+ :\left(\psi^\dag_R \psi_R\right)^2 :
&=& - \half\left[ \left(\p_\t - i\p_\s \right) \phi \left(\p_\t - i\p_\s \right) \phi + 
\left(\p_\t + i\p_\s \right) \phi \left(\p_\t + i\p_\s \right) \phi \right] \nn\\
&=& -\left(\p_\t \phi \p_\t \phi - \p_\s \phi \p_\s \phi \right) . \label{third}
\eeq
We should observe that the third term is asymmetric under the exchange operation $\t \leftrightarrow \s$, {\it
i.e.} the Fermi-Bose equivalence mapping does not preserve the chirality. Once the Fermi-Bose 
equivalence mapping is completed, the chirality should be redefined. 
Collecting the bosonized terms, Eqs.(\ref{free}, \ref{second}, \ref{third}), we have the bulk action of the 
boson theory 
\beq
S_{Bulk} &=&  \int_0^\infty d\t \int^{\pi}_{-\pi} d\s \,
\frac{1}{4\pi} \Bigl[\left(1+g_2 - g_4 \right) \p_\t \phi \p_\t \phi + 
\left(1+g_2 + g_4 \right) \p_\s \phi \p_\s \phi \Bigr] . 
\eeq
Since we do not want the interactions of the fermion theory to change the signature of the world sheet metric, 
we may impose some constraint condition for the coupling parameters
\beq
-g_2 -1 < g_4 < g_2 +1 . \label{domain}
\eeq
If the theory does not contain the boundary interaction with the impurity, the model is exactly solvable. 
It should be noted that we do not separate the Hamiltonian to apply the Fermi-Bose equivalence, since 
the Hamiltonian of the fermion theory is not directly transcribed into the Hamiltonian of the 
corresponding boson theory by the equivalence. We also note that that the action is not written in terms of 
the chiral boson fields, $\phi_L$ and $\phi_R$. Since the four fermi interaction terms do not respect the 
chirality, which is initially defined by the fermion theory, the bosonized action is rewritten in terms of the 
non-chiral boson field $\phi$. 

To make this point clear, let us take a look at the Fermi-Bose equivalence relations 
Eqs.(\ref{bilinear1}, \ref{bilinear2}). By the Fermi-Bose equivalence or the bosonization, the 
fermion bilinears $\psi^\dag_L \psi_L$ and $\psi^\dag_R \psi_R$ may be written as $\sqrt{2} \p_\s \phi_L$
and $\sqrt{2} \p_\s \phi_R$ respectively, but they can be also
written as $\sqrt{2} i \p_\t \phi_L$ and $-\sqrt{2} i \p_\t \phi_R$. Thus, even if the four fermi interaction terms appear in the Hamiltonian of the fermion theory, they should not be treated as part of the Hamiltonian
of the boson theory. It is more appropriate to express them in terms of the non-chiral boson field $\phi$ and 
to redefine its canonical momentum field. Accordingly the Hamiltonian and the chiral boson fields of the bosonized action must be redefined in the scheme of the canonical quantization. 

The boundary interaction term can be bosonized as a periodic potential term 
\beq
S_{Boundary} &=& \frac{1}{2\pi}  \int^{\pi}_{-\pi} d\s \,{\l} \left(\psi^\dag_L \psi_R + \psi^\dag_R \psi_L \right)\Bigl\vert_{\t= 0}\nn\\
&=& \frac{1}{2\pi}  \int^{\pi}_{-\pi} d\s \,{\l} \left(e^{i\sqrt{2}\phi} + e^{-i\sqrt{2}\phi} \right)\Bigl\vert_{\t= 0} . 
\eeq
Collecting the bosonized terms, which are all expressed in terms of the non-chiral boson field $\phi$, 
we find the action of the boson theory 
\beq
S = \frac{1}{4\pi} \int d\t d\s \Bigl[\left(1+g_2 - g_4 \right) \p_\t \phi \p_\t \phi + 
\left(1+g_2 + g_4 \right) \p_\s \phi \p_\s \phi \Bigr] + 
\frac{\l}{2\pi} \int d\t \left(e^{i\sqrt{2}\phi} + e^{-i\sqrt{2}\phi} \right)\Bigl\vert_{\t=0} . 
\eeq
Scaling the boson field $\phi$ as $\phi \rightarrow \frac{1}{\sqrt{2}} \phi$, 
\beq
S = \frac{1}{8\pi} \int d\t d\s \Bigl[\left(1+g_2 - g_4 \right) \p_\t \phi \p_\t \phi + 
\left(1+g_2 + g_4 \right) \p_\s \phi \p_\s \phi \Bigr] + 
\frac{\l}{2\pi} \int d\t \left(e^{i\phi} + e^{-i\phi} \right)\Bigl\vert_{\t=0} . 
\eeq
It is equivalent to a finite renormalization of the wave functions. In oder to recast the theory into 
the canonical setting, where we can apply the Fermi-Bose equivalence, we 
scale $\t$ and $\s$, using a finite conformal transformation 
\beq
\t \rightarrow a \t, ~~~ \s \rightarrow b \s .
\eeq
It brings us to the following action
\beq
S = \frac{ab}{8\pi} \int d\t d\s \Bigl[\frac{\left(1+g_2 - g_4 \right)}{a^2} \p_\t \phi \p_\t \phi + 
\frac{\left(1+g_2 + g_4 \right)}{b^2}\p_\s \phi \p_\s \phi \Bigr] + 
\frac{\l a}{2\pi} \int d\t \left(e^{i\phi} + e^{-i\phi} \right)\Bigl\vert_{\t=0} .
\eeq 
If we choose 
\beq
\frac{a}{b} = \sqrt{\frac{1+g_2 - g_4 }{1+g_2 + g_4 }}, 
\eeq
we can recast the boson action into the canonical setting 
\beq
S = \frac{1 + g}{4\pi} \int d\t d\s \left[\p \phi \p \phi \right] + \frac{V_0}{2\pi} \int d\t  \left(e^{i\phi} + e^{-i\phi} \right)\Bigl\vert_{\t=0} ,
\eeq
where 
\beq \label{couplingg}
g = -1 + \half \sqrt{(1+g_2)^2 - g_4^2 }, ~~~ V_0 = \l \a .
\eeq
This is the well-known Schmid model, which has been discussed extensively in the literature. The TL parameter
in the literature may be identified as 
\beq
K = 1+g = \half \sqrt{(1+g_2)^2 - g_4^2 }.
\eeq

\begin{figure}[htbp]
   \begin {center}
    \epsfxsize=0.6\hsize
%
	\epsfbox{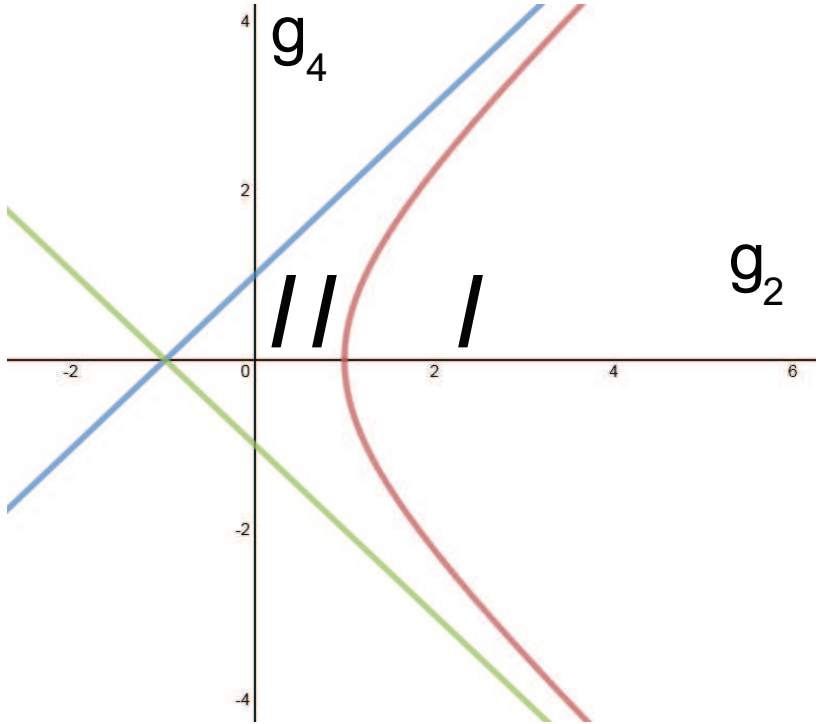}
   \end {center}
   \caption {\label{phase} Phase Diagram and Critical Line of Tomonaga-Luttinger Liquid}
\end{figure}

The critical line, which defines the phase boundary on the phase diagram of the theory follows from 
the renormalization group (RG) analysis of the coupling constant $V$, which has been performed 
in the previous works. The RG flow of $V$ is obtained \cite{schmid, LeeU(1)} as 
\beq
V^2 = V_0 \left(\frac{\Lambda^2}{\mu^2}\right)^{\frac{g}{2(g+1)}} .
\eeq
The physical domain is bounded by the two lines Eq.(\ref{domain}) : $g_4 = \pm (g_2 +1)$
and the critical line is the right branch of the hyperbola 
\beq
\frac{\left(1+g_2\right)^2}{2^2} - \frac{g^2_4}{2^2} = 1 .
\eeq
Where $g > 0$ the periodic potential becomes a relevant operator and where $g < 0$ the
potential becomes an irrelevant operator. In the region where $g > 0$, the periodic
potential tends to be strong and in the other region where $g < 0$, the potential tends to be weak 
in the low energy limit. 
Thus, in the region $I$ of the Fig.\ref{phase}, in the low energy limit (or the zero temperature limit),
the quantum wire behave two independent pieces of disconnected wire (total reflection) 
while in the region $II$ the quantum wire behaves as a single piece of connected wire (perfect transmission)
\cite{kane;1992b}.  

\section{Conclusions}

We studied the field theoretical model of the TLL and its critical behaviour, extending Coleman's work on 
the equivalence between the Thirring model and the sine-Gordon model. 
The model describes a quantum wire on a half line with a Dirac fermion 
in the folded setup. In the unfolded setup, the model is equivalent to the model of a single chiral fermion
on an infinite line with an impurity in the middle \cite{Nayak, Leechiral}. 
The traditional treatment of the TLL model is 
to apply the Fermi-Bose equivalence to the fermion Hamiltonian and to diagonalize the bosonized
Hamiltonian by the Bogoliubov transformation. In this paper we took an alternative route. The main tool of 
analysis was, of course the Fermi-Bose equivalence. But we recasted the fermion theory into the 
canonical setting before applying the Fermi-Bose equivalence, since the Fermi-Bose equivalence only works 
for certain values of the parameters of the theory. Then applying the equivalence to the action of the 
fermion model, we rewrite the bosonized action in terms of the non-chiral boson field $\phi$. 
The fermion bilinear operators are mapped into spatial derivatives or time derivatives of the chiral boson
fields. Although the two chiral boson fields commute with each other, they should not be treated as two 
completely independent boson fields with different flavors in the scheme of the canonical quantization. 
The notion of the chiral decomposition involves already a time derivative in the boson theory. 
Even if the four fermi interaction
terms appear in the Hamiltonian of the fermion theory, the bosonized four fermi interaction 
terms should not be treated as part of the Hamiltonian of the corresponding boson theory. 
They may be written as spatial derivatives of the chiral boson fields, but they can be also 
equally written as time derivatives of the chiral boson fields. Being time derivatives of the boson fields,
the bosonized four fermi interaction terms shift the canonical momentum boson field. It requires us to redefine 
the momentum field and the Hamiltonian. Thus, the Fermi-Bose equivalence does not preserve 
the chirality and the Hamiltonian. 

We also did not make use of the Bogoliubov transformation, which mixes the two chiral boson fields, 
$\phi_L$ and $\phi_R$ to diagonalize the Hamiltonian. The Bogoliubov
transformation formally diagonalizes the Hamiltonian, but at the same time it alters 
the boundary interaction terms and the 
boundary condition. The Bogoliubov transformation, which mixes the two chiral boson fields, may not be useful when we deal with a theory defined on the space-time with a boundary. It may help us to diagonalize the bulk Hamiltonian, but in return it makes the boundary condition and the boundary interaction complicated. 

In this paper we only make use of the transformations which preserve the symmetries of the system; the finite 
conformal transformation and the finite wave function renormalization. Bosonizing the fermion action by the 
Fermi-Bose equivalence, we rewrite the boson action in terms of the non-chiral boson fields instead of the 
chiral boson fields $\phi_L$ and $\phi_R$. Then the finite conformal transformation and the finite 
wave function renormalization recast the boson theory into the Schmid model. The chirality is seemingly 
broken by the Fermi-Bose equivalence mapping. But the theory itself redefines its own chirality. 
The chiral boson fields of the resultant theory do not correspond to the bilinears of the chiral fermion fields
precisely due to the four fermi interactions. The low energy behaviour of the theory critically depends 
the coupling $g$ Eq.(\ref{couplingg}). At $g=0$, the theory becomes exactly solvable and critical. In the region where 
$g >0$, the strength of the boundary periodic potential becomes strong in the low energy limit. In this region
kinks, which are soliton solutions to the equation of motion, appear as new dynamical degrees of freedom. 
The dual boson field, which describes the dynamics of the multi-kink configurations, replaces the boson field
$\phi$ in this phase. The effective action for the dual field takes the same form of the Schmid model but 
with coupling constant inverted:  $\a = g+1 \rightarrow 1/\a$. This is called the particle-kink duality \cite{schmid}. 
Under this duality the strong coupling phase is mapped onto the weak coupling phase. It is interesting 
to examine the realization of the particle-kink duality in the TLL model. We may further extend the 
present work to the spin-dependent Tomonaga-Luttinger model \cite{Furusaki93, LeeU(1)}. 
It may be also worthwhile to 
study the effects of the four fermi interactions in the Kondo problem \cite{Affleck:1990by, Affleck:1990iv, 
Furusaki94} and the quantum Brownian motion on
a triangular lattice \cite{Lee2009q} along the line of extension of this work.

\vskip 1cm

\begin{acknowledgments}
This work was supported by Kangwon National University Research Grant 2013.
\end{acknowledgments}


\end{document}